\documentclass[a4paper,11pt]{article}
\pdfoutput=1 

\usepackage{jheppub}

\usepackage{graphicx, epstopdf}

\usepackage[utf8]{inputenc}

\usepackage{amsmath, amsthm, amsfonts, amssymb}
\usepackage{mathtools, mathabx, bm, bbm, scalerel, xfrac}
\usepackage{empheq, physics, slashed, cancel, braket}

\usepackage{hyperref}
\usepackage{color}

\def\bT{\boldsymbol{T}}
\def\msbar{\overline{\text{MS}}}

\allowdisplaybreaks

\title{How much color do we really need? \\
Two-loop subleading-color effects in photon and jet physics}

\author[a]{Micha\l{} Czakon,}
\author[b]{Rene Poncelet}

\affiliation[a]{Institut f\"ur Theoretische Teilchenphysik und Kosmologie, RWTH Aachen University,\\ 52056 Aachen, Germany}
\affiliation[c]{The Henryk Niewodnicza\'{n}ski Institute of Nuclear Physics,\\ ul.\ Radzikowskiego 152, 31-342 Krakow, Poland}

\emailAdd{mczakon@physik.rwth-aachen.de}
\emailAdd{rene.poncelet@ifj.edu.pl}

\abstract{In recent years, the complete set of cross sections for Large Hadron Collider (LHC) processes ending with three resolved final states consisting of either photons or jets has been evaluated at next-to-next-to-leading order in QCD and leading order in QED. Results for three photons or three jets have only been obtained using the leading-color approximation of the virtual two-loop amplitudes. In the meantime, the required amplitudes have become available without recourse to the color expansion. In the present publication, we quantify the effects of the subleading-color contributions, and show that they do not exceed 2\% for most of the previously published results. The one exception is the ratio of three- to two-jet cross sections, where subleading-color effects can reach up to 5\%. Furthermore, we show that these conclusions hold for both popular infrared renormalization schemes, minimal subtraction and Catani's. The size of the effects is usually overshadowed by the size of the remaining uncertainty due to the truncation of the perturbation series. This is particularly important in the case of three-jet distributions that have already been used for the extraction of the strong-coupling constant at the very high energies available at the LHC.}
\preprint{IFJPAN-IV-2025-24, P3H-25-115, TTK-25-46}

\begin{document}
\maketitle
\flushbottom

\section{Introduction}

The expansion of SU($N_c$)-gauge-theory correlation functions in the number of colors, $N_c$, has been recognized as a valuable tool in the seminal papers of 't Hooft \cite{tHooft:1973alw} and Witten \cite{Witten:1979kh}. The leading term of the expansion is exclusively given by planar diagrams, which allows for a qualitative analysis of interactions even in the non-perturbative regime. Planar diagrams are not only a potentially small subset of all diagrams occurring in a given problem, but also have a simpler structure and evaluate to simpler functions. These facts are particularly advantageous in the case of virtual corrections to scattering cross sections.

In the present publication, we shall consider interactions between quarks and gluons at high orders of perturbative QCD, and include the possibility of emission of real photons from quark lines. Unfortunately, the presence of photons complicates the evaluation of Feynman diagrams, since a given planar QCD diagram may become non-planar once a photon is emitted from an internal line. Still, even in this case, computations are easier to perform in the leading-color approximation. In the worst case, one can always neglect the non-planar diagrams, as long as they form a gauge-invariant subset. 

Several results have been obtained for processes of the specified class, and we restrict ourselves to hadron-collision cross sections resulting in three final-state objects, photons and/or jets, at next-to-next-to-leading order (NNLO). Befitting our purpose, the required two-loop amplitudes are known with complete $N_c$ dependence for five partons \cite{Badger:2019djh,Agarwal:2023suw,DeLaurentis:2023nss,DeLaurentis:2023izi}, four partons and a photon \cite{Badger:2023mgf}, three partons and two photons \cite{Agarwal:2021vdh,Badger:2021imn}, and a quark pair and three photons \cite{Abreu:2023bdp}. These results superseded leading-color approximations provided in Refs.~\cite{Gehrmann:2015bfy,Badger:2018enw,Abreu:2018zmy,Abreu:2019odu,Abreu:2021oya} for five partons, in Refs.~\cite{Agarwal:2021grm,Chawdhry:2021mkw} for a quark pair, a gluon and two photons, and in Refs.~\cite{Chawdhry:2020for,Abreu:2020cwb} for a quark pair and three photons. Interestingly, the amplitudes for four partons and a photon have been obtained directly in full color, demonstrating the quick advances of the field.

Theoretical cross-section predictions for the selected processes based on full-color amplitudes have only been published for $pp \to \gamma + 2$~jets \cite{Badger:2023mgf}, and $pp \to \gamma \gamma +$~jet \cite{Buccioni:2025bkl}, the latter superseding the leading color studies of Refs.~\cite{Chawdhry:2021hkp,Badger:2021ohm}.
Our immediate purpose is, therefore, to determine the impact of the subleading-color contributions on the cross sections for $pp \to 3$ jets by comparing to Ref.~\cite{Alvarez:2023fhi}, and on the cross sections for $pp \to \gamma\gamma\gamma$ by comparing to Ref.~\cite{Chawdhry:2019bji} (see also Ref.~\cite{Kallweit:2020gcp}). Even though results for three photons in the final state have been obtained earliest in the considered class of processes, they have still not been upgraded to full color. The impact of subleading-color contributions has only been roughly estimated in Ref.~\cite{Abreu:2023bdp}. As far as $pp \to \gamma + 2$~jets is concerned, subleading-color contributions impact the result at the 2\% level according to Ref.~\cite{Badger:2023mgf}. The recent publication \cite{Buccioni:2025bkl} discusses subleading-color contributions in $pp \to \gamma\gamma + $~jet using a different definition to ours, as we shall explain shortly. Hence, we will also make statements on this process by comparing to Ref.~\cite{Chawdhry:2021hkp}. Recently, another approximate theory prediction \cite{Czakon:2021mjy} for $pp \to 3$~jets has been provided for comparison to the ATLAS collaboration \cite{ATLAS:2024png}. We will assess the impact of subleading-color contributions in this case as well.

Although an expansion in $N_c$ is well defined at the level of partonic cross sections, it would not be wise to take its leading term as a good approximation of the complete result. Subleading-color contributions are formally of $\order{1/N_c^2}$, i.e.\ 10\% for QCD, but order-of-magnitude estimates are only reliable up to numerical factors. Hence, the leading-color approximation to a cross section may be far more than 10\% away from the actual result. It has, therefore, become the practice to only take the leading-color approximation of the two-loop virtual corrections and, sometimes, of the square of the one-loop virtual corrections. This approximation is subsequently rescaled by the ratio of the full-color and leading-color Born amplitudes. The virtual corrections are divergent due to the presence of massless partons as external states. These divergences of soft and/or collinear origin are well understood by now \cite{Catani:1998bh,Dixon:2008gr,Gardi:2009qi,Becher:2009cu,Becher:2009qa} and allow to define a meaningful finite part of the virtual amplitudes. It is only this finite part, the finite remainder, that is approximated in leading color. This introduces a dependence on the renormalization scale used to evaluate the leading-color finite remainder. Furthermore, one can define different finite remainders by absorbing finite contributions into the divergences. There are currently two infrared-renormalization schemes used for this purpose, minimal subtraction \cite{Becher:2009cu} and Catani's \cite{Catani:1998bh}. As in our previous studies, we will use minimally subtracted two-loop amplitudes in dimensional regularization, but we will also quantify the consequences of using Catani's formula.

It is to be expected that many future calculations of cross sections will still use the leading-color approximation, as has been done for example very recently in Refs.~\cite{Badger:2025ilt,Badger:2025ljy}. The quality of this approach can only be ascertained once full-color results become available. Still, it is useful to have case studies that would increase the confidence in results derived in this way. This is our main motivation for the present publication.

This paper is organized as follows. In the next section, we discuss the definition of the two-loop finite remainder, which is the object that has been approximated in previous calculations. Afterwards, we evaluate the impact of subleading-color contributions in published results. To illustrate our conclusions, we only present selected differential distributions. The complete set of plots corresponding to the referenced publications is included with the arXiv submission of this work. We also verify that a nuisance parameter as in Ref.~\cite{Lim:2024nsk} can be used to obtain a reliable estimate of the size of the subleading color contributions. This topic is discussed in the last section right before our conclusions.

\section{The ambiguous finite remainder}

As explained in the introduction, the leading-color expansion is applied to the finite remainder, $|{\cal F}^{(\text{S})}(\{\underline{p}\},\mu)\rangle$, of a scattering amplitude $|{\cal M}(\epsilon,\{\underline{p}\})\rangle$ obtained in dimensional regularization with spacetime dimension $d \equiv 4-2\epsilon$,
\begin{equation}\label{renorm}
   |{\cal F}^{(\text{S})}(\{p_i\},\mu)\rangle 
   \equiv \lim_{\epsilon\to 0}\,
   \Big[ \bm{Z}^{(\text{S})}(\epsilon,\{p_i\},\mu) \Big]^{-1}\,
   |{\cal M}(\epsilon,\{p_i\})\rangle \,.
\end{equation}
Both the finite remainder and the original amplitude are vectors in color and spin space, and depend on the kinematics, $\{p_i\}$. As usual, $\mu$ denotes the scale parameter of dimensional regularization, while the superscript $(\text{S})$ specifies the scheme in which the finite remainder has been defined. The finite remainder can be expanded perturbatively as follows
\begin{equation}
   |{\cal F}^{(\text{S})}(\mu) \rangle \equiv \alpha_s^m(\mu) \bigg[ |{\cal F}^{(\text{S},0)}(\mu) \rangle + \frac{\alpha_s(\mu)}{4\pi}\, |{\cal F}^{(\text{S},1)}(\mu) \rangle + \left( \frac{\alpha_s(\mu)}{4\pi} \right)^2 |{\cal F}^{(\text{S},2)}(\mu) \rangle + \dots \bigg] \,,
\end{equation}
where $m$ is the power of the strong coupling constant $\alpha_s$ in the Born approximation, and
\begin{equation}
\alpha_s^m(\mu) \, |{\cal F}^{(\text{S},0)}(\mu) \rangle = \eval{ |{\cal M}(\epsilon = 0)\rangle }_{\substack{\text{tree} \\ \text{level}}} \equiv \alpha_s^m(\mu) \, |{\cal M}^{(0)} \rangle \,.
\end{equation}

Two schemes are currently widely used. The first one is due to the work of Catani \cite{Catani:1998bh} who was the first to propose a formula for the divergences of an arbitrary two-loop scattering amplitude involving massless external partons
\begin{equation}
   \Big[ \bm{Z}^{(\rm{Catani})} \Big]^{-1} \equiv 1 - \frac{\alpha_s}{2\pi}\,\bm{I}^{(1)}
   - \left( \frac{\alpha_s}{2\pi} \right)^2 \bm{I}^{(2)}
   + \dots \,,
\end{equation}
where the subtraction operators are
\begin{eqnarray}\label{I2}
\begin{aligned}
   \bm{I}^{(1)}(\epsilon) 
   &= \frac{e^{\epsilon\gamma_E}}{\Gamma(1-\epsilon)}\,
    \sum_i \left( \frac{1}{\epsilon^2} 
    - \frac{\gamma_0^i}{2\epsilon}\,\frac{1}{\bm{T}_i^2} \right)
    \sum_{j\neq i}\,\frac{\bm{T}_i\cdot\bm{T}_j}{2}
    \left( \frac{\mu^2}{-s_{ij}} \right)^\epsilon , \\[.2cm]
   \bm{I}^{(2)} 
   &= \frac{e^{-\epsilon\gamma_E}\,\Gamma(1-2\epsilon)}%
          {\Gamma(1-\epsilon)} 
    \left( \frac{\gamma_1^{\rm cusp}}{8}
     + \frac{\beta_0}{2\epsilon} \right) 
    \bm{I}^{(1)}(2\epsilon) 
    - \frac12\,\bm{I}^{(1)}(\epsilon)  
    \left( \bm{I}^{(1)}(\epsilon) + \frac{\beta_0}{\epsilon} \right)  
    + \bm{H}^{(2)} \,.
\end{aligned}
\end{eqnarray}
The sum in the first expression runs over partons, and $s_{ij}\equiv 2\sigma_{ij}\,p_i\cdot p_j+i0$, with $\sigma_{ij}=+1$ if the momenta $p_i$ and $p_j$ are both incoming or outgoing, and $\sigma_{ij}=-1$ otherwise. The quantity $\bm{H}^{(2)}$ was not determined in complete generality in the original publication. However, it is now known to be \cite{Aybat:2006mz,Becher:2009cu,Becher:2009qa}
\begin{equation}\label{H2}
\begin{split}
   \bm{H}^{(2)} 
   = \frac{e^{\epsilon \gamma_E}}{4\epsilon\Gamma(1-\epsilon)}  \bigg(\frac{\mu^2}{Q^2}\bigg)^{2\epsilon} &\Bigg[ \frac{1}{4}\,\sum_i \bigg( \gamma_1^i 
    - \frac14\,\gamma_1^{\rm cusp}\,\gamma_0^i
    + \frac{\pi^2}{16}\,\beta_0\,\gamma_0^{\rm cusp}\,C_i \bigg) \\
   &+ \frac{if^{abc}}{6}\,
   \sum_{(i,j,k)} \bm{T}_i^a\,\bm{T}_j^b\,\bm{T}_k^c\,
    \ln\frac{-s_{ij}}{-s_{jk}} \ln\frac{-s_{jk}}{-s_{ki}} 
    \ln\frac{-s_{ki}}{-s_{ij}}  \\
     &- \frac{if^{abc}}{32} \,\gamma_0^\text{cusp}
\sum_{(i,\,j,\,k)}\bT_i^a\,\bT_j^b\,\bT_k^c\;
\bigg( \frac{\gamma_0^i}{C_i} - \frac{\gamma_0^j}{C_j} \bigg)
\ln\frac{-s_{ij}}{-s_{jk}}\,\ln\frac{-s_{ki}}{-s_{ij}} \Bigg] \,.
\end{split}
\end{equation}
In Refs.~\cite{Catani:1998bh,deFlorian:2012za} which concern the Drell-Yan process and Higgs-boson production in gluon fusion respectively, $Q^2 = -s$ with $s$ the center-of-mass energy squared. On the other hand, in Refs.~\cite{Agarwal:2021vdh,Abreu:2023bdp,Abreu:2020cwb,Abreu:2018zmy,Abreu:2019odu,Abreu:2021oya,Agarwal:2021grm} that provide finite remainders in Catani's scheme for processes relevant to the present publication, $Q^2 = \mu^2$. The second and third lines in Eq.~\eqref{H2} vanish at leading color and for processes with at most three partons. Hence, this general form has not been used in any of the publications cited here, as the first line was always sufficient. The summation index in the second and third lines, $(i,j,k)$, denotes unordered tuples of distinct parton indices, whereas $C_i$ stands for the Casimirs $C_F$ and $C_A$ for quarks and gluons respectively. The listed formulae involve several anomalous-dimension coefficients, which we reproduce for completeness
\begin{eqnarray}
   \gamma^{\rm cusp} &=& \sum_{n=0}^\infty \gamma_n^{\rm cusp} \left( \frac{\alpha_s}{4\pi} \right)^{n+1}\,, \qquad \gamma^{q,g} = \sum_{n=0}^\infty \gamma_n^{q,g} \left( \frac{\alpha_s}{4\pi} \right)^{n+1}\,, \nonumber\\[0.2cm]
   \gamma_0^{\rm cusp} &=& 4 \,, \nonumber\\
   \gamma_1^{\rm cusp} &=& \left( \frac{268}{9} 
    - \frac{4\pi^2}{3} \right) C_A - \frac{80}{9}\,T_F n_f \,, \nonumber \\[.2cm]
      \gamma_0^q &=& -3 C_F \,, \nonumber\\
   \gamma_1^q &=& C_F^2 \left( -\frac{3}{2} + 2\pi^2
    - 24\zeta_3 \right)
    + C_F C_A \left( - \frac{961}{54} - \frac{11\pi^2}{6} 
    + 26\zeta_3 \right)
    + C_F T_F n_f \left( \frac{130}{27} + \frac{2\pi^2}{3} \right) \,, \nonumber \\[.2cm]
   \gamma_0^g &=& - \beta_0 
    = - \frac{11}{3}\,C_A + \frac43\,T_F n_f \,, \nonumber\\
   \gamma_1^g &=& C_A^2 \left( -\frac{692}{27} + \frac{11\pi^2}{18}
    + 2\zeta_3 \right) 
    + C_A T_F n_f \left( \frac{256}{27} - \frac{2\pi^2}{9} \right)
    + 4 C_F T_F n_f \,.
\end{eqnarray}
As usual, $n_f$ denotes the number of massless quark flavors.

The results for cross sections presented in Refs.~\cite{Chawdhry:2019bji,Chawdhry:2021hkp,Badger:2023mgf,Alvarez:2023fhi,Czakon:2021mjy} have been obtained with the sector-improved residue subtraction scheme constructed according to Refs.~\cite{Czakon:2010td,Czakon:2019tmo,Czakon:2021mjy}. Part of this construction is the definition of the two-loop finite remainder through minimal subtraction \cite{Becher:2009cu}. In this case
\begin{equation}
   \bm{Z}^{(\msbar)} \equiv 1 + \frac{\alpha_s}{4\pi}\,\bm{Z}_1
   + \left( \frac{\alpha_s}{4\pi} \right)^2 \bm{Z}_2
   + \dots \,,
\end{equation}
with the Laurent series of $\bm{Z}_{1,2}$ in $\epsilon$ given exclusively by the principal part and related to Catani's formula by
\begin{equation}
   \bm{Z}_1 = 2\bm{I}^{(1)} \Big|_{\substack{\text{principal} \\ \text{part}}} \,,
    \qquad
   \bm{Z}_2 = \Big[ 4\bm{I}^{(2)} + 2\bm{I}^{(1)} \bm{Z}_1 \Big]_{\substack{\text{principal} \\ \text{part}}} \,.
\end{equation}
$\bm{Z}^{(\msbar)}$ has an anomalous dimension of well-understood structure
\begin{equation} \label{eq:ScaleOfMS}
    \dv{\ln \mu} \ln \bm{Z}^{(\msbar)} = -\gamma^{\rm cusp} \sum_{i\neq j}\,\frac{\bm{T}_i\cdot\bm{T}_j}{2} \ln( \frac{\mu^2}{-s_{ij}} ) - \sum_i \gamma^i + \order{\alpha_s^3} \,.
\end{equation}
On the other hand, $\bm{Z}^{(\text{Catani})}$ is scale independent up to NNLO with $Q$ independent of $\mu$ in Eq.~\eqref{H2}
\begin{equation} \label{eq:ScaleOfCatani}
    \dv{\ln \mu} \bm{Z}^{(\text{Catani})} = \order{\alpha_s^3} \,.
\end{equation}
If instead $Q = \mu$, then the right hand side of Eq.~\eqref{eq:ScaleOfCatani} is of $\order{\alpha_s^2}$ but still free of $\ln(\mu)$.

In consequence of Eq.~\eqref{eq:ScaleOfCatani}, the two-loop finite remainder obtained with Catani's formula contains explicit logarithms of $\mu$ only up to $\ln^2(\mu)$
\begin{equation} \label{eq:ScaleOfCataniF}
|{\cal F}^{(\text{Catani},2)}(\mu) \rangle = |{\cal F}^{(\text{Catani},2)}(\mu_0) \rangle + \sum_{i=1}^2 \sum_{i \leq j \leq 2} c_{ij} \, \ln^i\Big( \frac{\mu}{\mu_0}\Big) \, |{\cal F}^{(\text{Catani},2-j)}(\mu_0) \rangle \,.
\end{equation}
On the other hand, Eq.~\eqref{eq:ScaleOfMS} implies that the $\msbar$ finite remainder has a stronger dependence on the scale up to $\ln^4(\mu)$
\begin{equation} \label{eq:ScaleOfMSF}
|{\cal F}^{(\msbar,2)}(\mu) \rangle = |{\cal F}^{(\msbar,2)}(\mu_0) \rangle + \sum_{i=1}^4 \sum_{i/2 \leq j \leq 2} d_{ij} \, \ln^i\Big( \frac{\mu}{\mu_0}\Big) \, |{\cal F}^{(\msbar,2-j)}(\mu_0) \rangle \,.
\end{equation}
$c_{ij}$ and $d_{ij}$ contain color operators acting on tree-level and one-loop matrix elements. The resulting contributions have been kept exact in $N_c$ in previous calculations. As we will see below, the leading color approximation is applied after factoring-out the tree-level matrix element and setting $\mu_0 = \hat{E}_{\text{CMS}}$. It is also often applied to the one-loop squared matrix element that enters the double-virtual cross section contribution, $\hat{\sigma}^{\mathrm{VVF}}$, to the NNLO partonic cross section,
\begin{equation}
    \hat{\sigma}^{(2)} \, \supset \, \hat{\sigma}^{\mathrm{VVF}} =
\frac{1}{2\hat{s}} \frac{1}{N} \int \mathrm{d} \bm{\Phi}_n \, \Big( 2
\Re \, \langle \mathcal{M}_n^{(0)} | \mathcal{F}_n^{(2)} \rangle + \langle \mathcal{F}_n^{(1)} |
\mathcal{F}_n^{(1)} \rangle \Big) \, \mathrm{F}_n \; .
\end{equation}
Here, $2\hat{s}$ is the partonic flux, $N$ is the number of partonic color and spin configurations, and $\mathrm{F}_n$ is the measurement function that defines the observable that is being evaluated.

\section{Subleading color in past results}

\subsection*{$pp \to \gamma\gamma\gamma$}

In Ref.~\cite{Chawdhry:2019bji}, the leading-color approximation has been applied to the two-loop finite remainder of the amplitude for the process $q\bar{q} \to \gamma\gamma\gamma$ \cite{Chawdhry:2020for,Abreu:2020cwb} at the scale equal to the partonic center-of-mass energy\footnote{The $n_f N_c$ term has been provided in Ref.~\cite{Chawdhry:2020for} as well, even though it has not been used for the cross section calculation.},
\begin{equation} \label{eq:ppaaa_fin2}
\begin{split}
    &\frac{ 2 \, \text{Re} \, \ip*{{\cal M}^{(0)}}{{\cal F}^{(\msbar,2)}(\mu = \hat{E}_{\text{CMS}})} }{\ip*{{\cal M}^{(0)}}{{\cal M}^{(0)}}}\\ &\qquad= N_c^2 \bigg[ c_{1,0} + \frac{1}{N_c^2} c_{1,1} + \frac{1}{N_c^4} c_{1,2} + \frac{n_f}{N_c} \Big( c_{2,0} + \frac{1}{N_c^2} c_{2,1} \Big) \bigg] + \sum_{q'} \frac{Q_{q'}^2}{Q_q^2} \, N_c \, \Big( c_{3,0} + \frac{1}{N_c^2} c_{3,1} \Big) \\[0.2cm]
    &\qquad\approx N_c^2 \, c_{1,0} \,,
\end{split}
\end{equation}
where $Q_q$ is the charge of the external quarks, while $Q_{q'}$ is the charge of quarks belonging to a loop of internal lines. The calculation employed dynamic, i.e.\ kinematics dependent, scales as specified in Appendix~\ref{app:setups}. The scale dependence was restored according to Eq.~\eqref{eq:ScaleOfMSF} without any further approximation. As in the experimental analysis \cite{ATLAS:2017lpx}, the following observables were studied: $p_T(\gamma_i)$ (photon transverse momentum), $\Delta \phi(\gamma_i,\gamma_j)$ (azimuthal-angle difference between photon momenta), $|\Delta \eta(\gamma_i, \gamma_j)|$ (rapidity difference between photon momenta), $m(\gamma_i,\gamma_j)$ (two-photon invariant mass), $m(\gamma_1,\gamma_2,\gamma_3)$ (three-photon invariant mass). 

\begin{figure}[t]
\includegraphics[width = 0.32\textwidth,page=1]{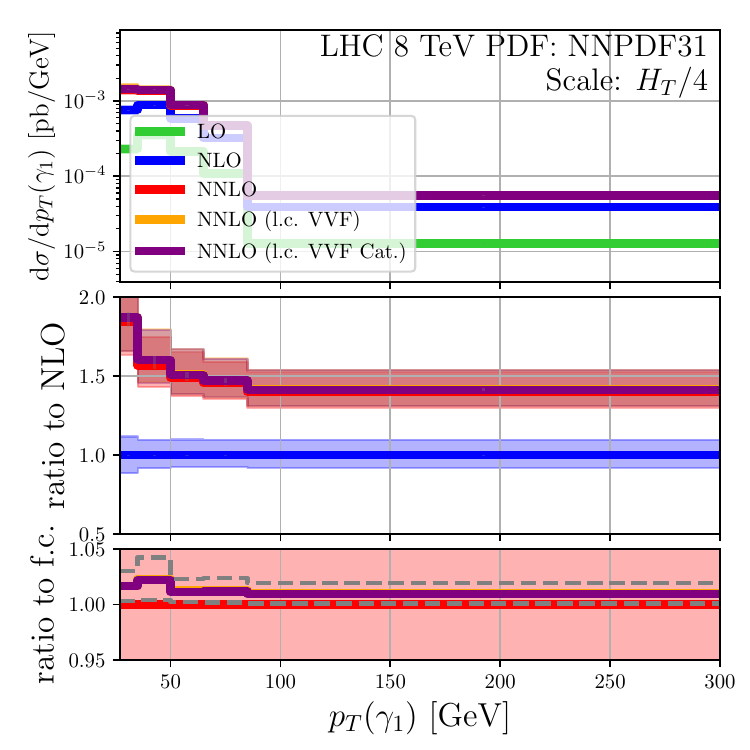}
\includegraphics[width = 0.32\textwidth,page=4]{figures/final_pp_aaa_1911.00479_NNPDF31_HTo4.pdf}
\includegraphics[width = 0.32\textwidth,page=7]{figures/final_pp_aaa_1911.00479_NNPDF31_HTo4.pdf}
\caption{Differential distributions for the $pp \to \gamma\gamma\gamma$ process. The lowest panel presents the ratio of the leading-color approximated results to the full-color prediction. Scale-uncertainty bands correspond to seven-point scale variation. Histograms "l.c. VVF Cat." have been obtained with Catani's finite remainder at leading color, while "l.c. VVF" with minimal subtraction. The dashed grey lines in the lowest panel demark the uncertainty estimated according to Section~\ref{sec:TNP}.}
\label{fig:aaa}
\end{figure}

As pointed out in Ref.~\cite{Chawdhry:2019bji}, the NNLO corrections are large for this process, but the leading-color $\msbar$ finite-reminder contribution does not exceed a few percent. As expected, therefore, the subleading-color contributions are small and, in fact, do not exceed 2\% for any of the studied distributions. We provide selected examples in Fig.~\ref{fig:aaa}, for which we used the full-color amplitudes from Ref.~\cite{Abreu:2023bdp}. It should be stressed that the leading color approximation is a rather poor approximation of the finite remainder itself for this process. On the other hand, both definitions of the finite remainder at leading color, $\msbar$ and Catani's, yield essentially the same results.

\subsection*{$pp \to \gamma\gamma+$jet}

\begin{figure}[t]
\includegraphics[width = 0.32\textwidth,page=3]{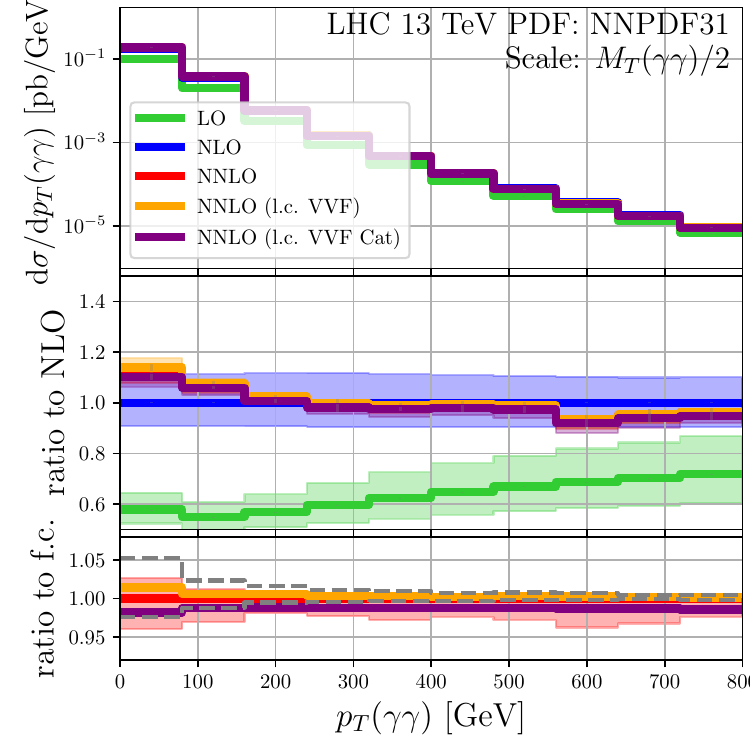}
\includegraphics[width = 0.32\textwidth,page=6]{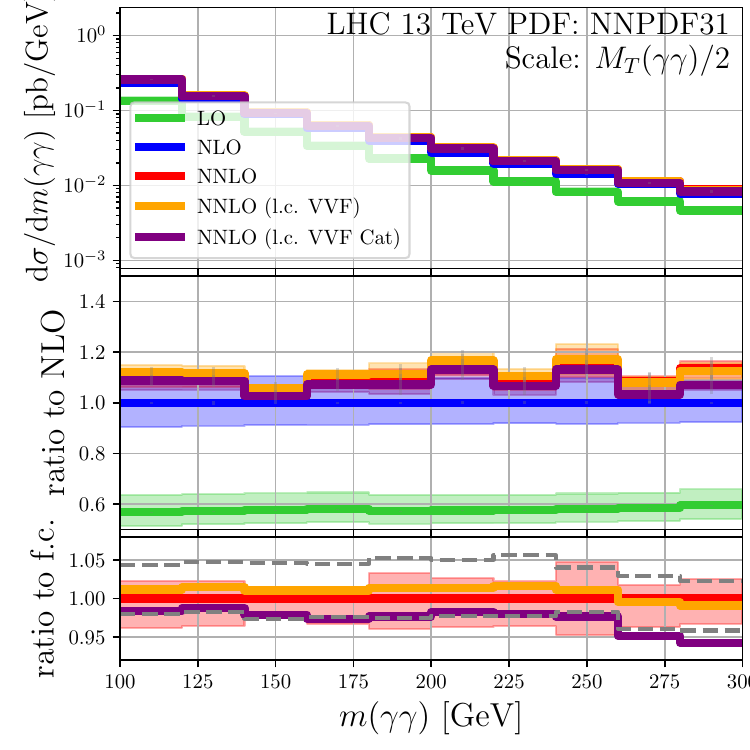}
\includegraphics[width = 0.32\textwidth,page=4]{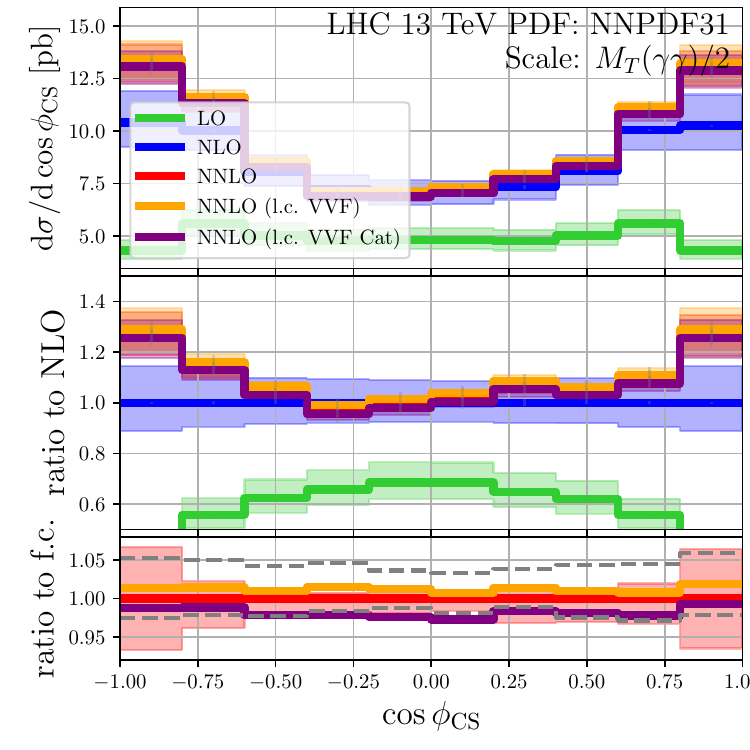}
\caption{Differential distributions for the $pp \to \gamma\gamma+$jet process. The format is the same as in Fig.~\ref{fig:aaa}.}
\label{fig:aaj}
\end{figure}

In Ref.~\cite{Chawdhry:2021hkp}, the leading-color approximation has been applied to the two-loop finite remainder of the amplitude for the process $q\bar{q} \to \gamma\gamma g$ \cite{Chawdhry:2021mkw,Agarwal:2021grm} at the scale equal to the partonic center-of-mass energy,
\begin{equation}\label{eq:ppaaj_fin2}
\begin{split}
    &\frac{ 2 \, \text{Re} \, \ip*{{\cal M}^{(0)}}{{\cal F}^{(\msbar,2)}(\mu = \hat{E}_{\text{CMS}})} }{\ip*{{\cal M}^{(0)}}{{\cal M}^{(0)}}} \\[0.2cm] &\qquad\qquad\qquad\qquad = N_c^2 \bigg[ c_{1,0} + \frac{1}{N_c^2} c_{1,1} + \frac{1}{N_c^4} c_{1,2} + \frac{n_f}{N_c} \Big( c_{2,0} + \frac{1}{N_c^2} c_{2,1} \Big) \bigg] \\[.2cm] &\qquad\qquad\qquad\qquad\quad + \sum_{q'} \frac{Q_{q'}}{Q_q} \, N_c \Big( c_{3,0} + \frac{1}{N_c^2} c_{3,1} \Big) + \sum_{q'} \frac{Q_{q'}^2}{Q_q^2} \, N_c \Big( c_{4,0} + \frac{1}{N_c^2} c_{4,1} + \frac{n_f}{N_c} \, c_{5,0} \Big) \Big) \bigg] \\[0.2cm]
    &\qquad\qquad\qquad\qquad \approx N_c^2 \bigg[ c_{1,0} + \frac{n_f}{N_c} c_{2,0} \bigg] + \sum_{q'} \frac{Q_{q'}^2}{Q_q^2} \, n_f \, c_{5,0} \,.
\end{split}
\end{equation}
In this case, it was possible to keep some contributions enhanced by $n_f$ given by planar diagrams only. The scale dependence was again restored with full $N_c$ dependence with Eq.~\eqref{eq:ScaleOfMSF}. The setup for the study is given in Appendix~\ref{app:setups}. Results were obtained for kinematic distributions of the photons, in particular for the pair transverse momentum $p_T(\gamma\gamma)$, rapidity $|\eta(\gamma\gamma)|$, and invariant mass $m(\gamma\gamma)$, as well as for the difference in azimuthal angle $\Delta\phi(\gamma\gamma)$, and in rapidity $\Delta \eta(\gamma\gamma)$. Additionally, distributions were calculated for the cosine of the angle between the two photons in the Collins-Soper frame \cite{Collins:1977iv}\footnote{See Eq. (7) in Ref.~\cite{ATLAS:2015zhl}}
\begin{equation}
    \cos \phi_{\text{CS}} \equiv \frac{\sinh(\Delta \eta(\gamma\gamma))}{\sqrt{1+\bigg( \dfrac{p_T(\gamma\gamma)}{m(\gamma\gamma)} \bigg)^2}} \frac{2 p_T(\gamma_1) p_T(\gamma_2)}{m^2(\gamma\gamma)} \,.
\end{equation}
The NNLO corrections are not as large for this process as for the production of three photons. Furthermore, the two-loop virtual contribution does not exceed 2\% for any of the distributions. Hence, it is to be expected that the difference between leading-color and full-color calculations will not be large either when compared to the complete prediction. This is indeed the case, and the effects of subleading-color contributions fall within the 2\% window. Some examples are provided in Fig.~\ref{fig:aaj}, for which we have used the amplitudes from Refs.~\cite{Agarwal:2021vdh,Badger:2021imn}. Compared to the full-color two-loop finite remainder, the leading-color approximation is again rather poor. At the same time, Catani's leading-color finite remainder seems to perform slightly worse than the $\msbar$ one.

Recently, another theoretical analysis for the diphoton+jet process has been published \cite{Buccioni:2025bkl} matching the experimental measurements from Ref.~\cite{ATLAS:2021mbt}. The setup is slightly different to that of Ref.~\cite{Chawdhry:2021hkp}. More importantly, however, the new calculation uses Catani's finite remainders and contains a comparison between full-color and leading-color approximations. From Fig.~2 of Ref.~\cite{Buccioni:2025bkl} it follows that subleading-color effects do not exceed 3\% for any of the distributions, and are typically under 1\% for kinematics away from singularities where resummation would be needed.

\subsection*{$pp \to \gamma+2$ jets}

For completeness, we summarize the findings of Ref.~\cite{Badger:2023mgf}, which contains a discussion of subleading-color contributions for the photon+di-jet production process. In that publication, the leading-color approximation has been applied not only to the two-loop finite remainder but also to the one-loop finite remainder squared for the two relevant processes $0 \to q\bar{q}gg\gamma$ and $0 \to q_1\bar{q}_1 q_2 \bar{q}_2 \gamma$. Thus,
\begin{equation} \label{eq:2M0F2+F1F1}
    \frac{ 2 \, \text{Re} \, \ip*{{\cal M}^{(0)}}{{\cal F}^{(\msbar,2)}(\mu = \hat{E}_{\text{CMS}})} + \ip*{{\cal F}^{(\msbar,1)}(\mu = \hat{E}_{\text{CMS}})}{{\cal F}^{(\msbar,1)}(\mu = \hat{E}_{\text{CMS}})}}{\ip*{{\cal M}^{(0)}}{{\cal M}^{(0)}}} \,,
\end{equation}
has been approximated by keeping only terms of $\order{N_c^2}$, $\order{n_f N_c}$ and $\order{\sum Q_{q'}/Q_{q} \, N_c}$. In consequence, some non-planar diagrams were included in the approximation. With these assumptions, the subleading-color contributions turned out to be completely negligible. This fact is partly due to the size of the double-virtual contributions, which did not exceed 10\% of the NNLO prediction for all distributions but that of the invariant mass of the two jets. Unfortunately, only the $\msbar$ leading-color finite remainder was tested, and we cannot quantify the effect of using Catani's leading-color finite remainders. On the other hand, it is to be expected that the subleading-color contributions would be just as negligible with this scheme.

\subsection*{$pp \to 3$ jets}\label{eq:ppjjj_fin2}

Theoretical predictions for three-jet production in hadron collisions at NNLO QCD have been published in Refs.~\cite{Alvarez:2023fhi,Czakon:2021mjy} (see also Ref.~\cite{Chen:2022ktf} for pure-gluon results). Here, we first discuss Ref.~\cite{Alvarez:2023fhi}, because it constituted the basis for a determination of the strong coupling constant at very high energies in Ref.~\cite{ATLAS:2023tgo}. Now that full-color two-loop virtual amplitudes are finally available \cite{Badger:2019djh,Agarwal:2023suw,DeLaurentis:2023nss,DeLaurentis:2023izi}, it is important to investigate whether the use of the leading-color approximation in Ref.~\cite{Alvarez:2023fhi} could have had an impact on the extracted value of $\alpha_s$.

The observable that we are interested in is the transverse energy-energy correlator (TEEC) at the LHC with 13 TeV center-of-mass energy
\begin{align}
  \frac{1}{\sigma_2}\frac{\dd \sigma}{\dd \cos\Delta\phi}
   \equiv \frac{1}{\sigma_2}
   \sum_{\text{jets } i \neq j}\int \frac{\dd \sigma\; x_{\perp,i} x_{\perp,j}}{\dd x_{\perp,i} \dd x_{\perp,j} \dd \cos\Delta\phi_{ij}}
    \delta(\cos\Delta\phi-\cos\Delta\phi_{ij}) \dd x_{\perp,i} \dd x_{\perp,j} \dd \cos\Delta\phi_{ij} \,,
\label{eq:TEEC}    
\end{align}
where $x_{\perp, i} = E_{\perp, i}/\sum_k E_{\perp,k}$, and $E_{\perp,i} = \sqrt{E_i^2-p_{z,i}^2}$ is the $i$-th jet transverse energy. The angle $\Delta\phi_{ij}$ is the azimuthal angle between jets $i$ and $j$. Jets are defined with the anti-$k_T$ algorithm with radius $R=0.4$, transverse momentum $p_T > 60$ GeV and rapidity $|y| < 2.4$. TEEC is measured in bins of
\begin{equation} \label{eq:HT2}
H_{T,2} = p_{T,1} + p_{T,2} \,, 
\end{equation}
where the transverse momenta are those of the two jets with highest $p_T$. In Eq.~\eqref{eq:TEEC}, $\sigma_2$ is the two-jet cross section evaluated in the same $H_{T,2}$ bin as the rest of the expression. In order to obtain reliable perturbative predictions at fixed order, the range of $\Delta \phi$ is restricted, since the endpoints $\cos \Delta \phi = \pm 1$ require resummation of soft-collinear emissions.

The theoretical predictions of Ref.~\cite{Alvarez:2023fhi} where obtained for renormalisation and factorisation scales
\begin{align} \label{eq:ScaleHT}
    \mu_R = \mu_F = \hat{H}_T = \sum_{\text{parton } i} p_{T,i}\,.
\end{align}
The leading-color approximation has been applied \cite{Abreu:2019odu} to the sum \eqref{eq:2M0F2+F1F1} of the two-loop finite remainder and the one-loop squared finite remainder defined in the $\msbar$ scheme and normalized to the tree-level amplitude. The result is given by planar diagrams only and contains terms of $\order{N_c^2 \, (n_f/N_c)^n}$, $n=0,1,2$.

\begin{figure}[t]
\includegraphics[width = 0.32\textwidth,page=1]{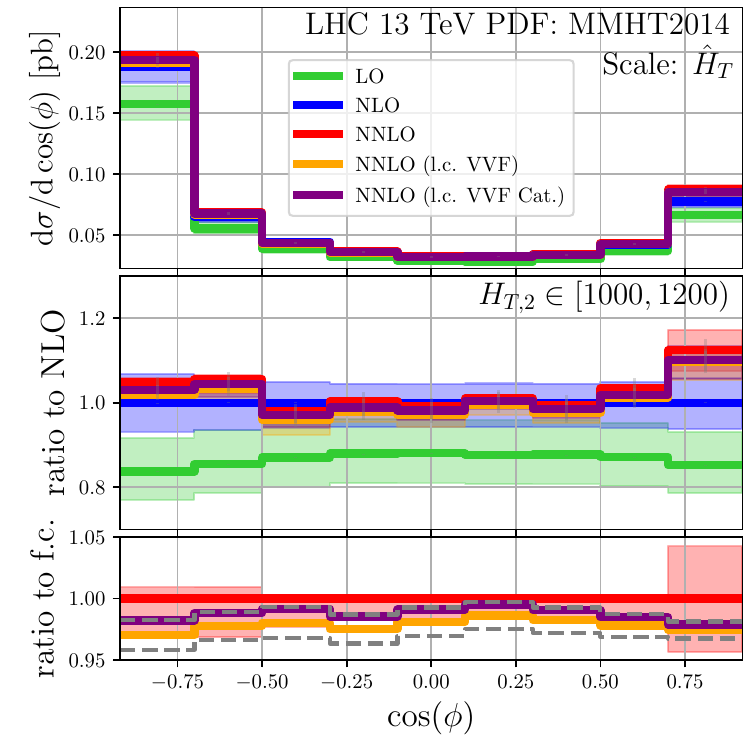}
\includegraphics[width = 0.32\textwidth,page=3]{figures/final_pp_jjj_2301.01086_MMHT2014_HThat}
\includegraphics[width = 0.32\textwidth,page=5]{figures/final_pp_jjj_2301.01086_MMHT2014_HThat}
\includegraphics[width = 0.32\textwidth,page=7]{figures/final_pp_jjj_2301.01086_MMHT2014_HThat}
\includegraphics[width = 0.32\textwidth,page=9]{figures/final_pp_jjj_2301.01086_MMHT2014_HThat}
\includegraphics[width = 0.32\textwidth,page=10]{figures/final_pp_jjj_2301.01086_MMHT2014_HThat}
\caption{Transverse energy-energy correlator. The format is the same as in Fig.~\ref{fig:aaa}.}
\label{fig:TEEC}
\end{figure}

In Fig.~\ref{fig:TEEC}, we provide results for TEEC obtained with full-color amplitudes in several $H_{T,2}$ bins obtained with the \textsc{MMHT2014} NNLO pdf set. The lowest panel of each plot reveals the size of the subleading-color contributions, which do not exceed 3\% in the first bin, and then become smaller and smaller with increasing $H_{T,2}$. Catani's leading-color finite remainder performs slightly better in this case. We note that the subleading-color contributions are covered by the scale uncertainty of the prediction. Hence, as long as the extraction of $\alpha_s$ accounts for scale uncertainties, the leading-color approximation is a viable method. Even more importantly, the purpose of Ref.~\cite{ATLAS:2023tgo} was to obtain $\alpha_s$ from the highest $H_{T,2}$ bins, where the leading-color approximation is barely different from the full-color result.

Let us now return to Ref.~\cite{Czakon:2021mjy}, which treats the ratio of three- and two-jet cross sections, $R_{3/2} \equiv \sigma_3/\sigma_2$. Just as TEEC, this ratio can be used for the determination of the strong coupling constant, as it has a nearly linear dependence on $\alpha_s$. At the time Ref.~\cite{Czakon:2021mjy} appeared, there were no measurements of $R_{3/2}$. Recently, however, the ATLAS collaboration has performed an analysis \cite{ATLAS:2024png} for which we have provided theoretical predictions based on leading-color amplitudes. Instead of updating Ref.~\cite{Czakon:2021mjy}, we will therefore adopt the setup of Ref.~\cite{ATLAS:2024png} and analyse the size of the subleading-color contributions in this context.

\begin{figure}[t]
\includegraphics[width = 0.32\textwidth,page=1]{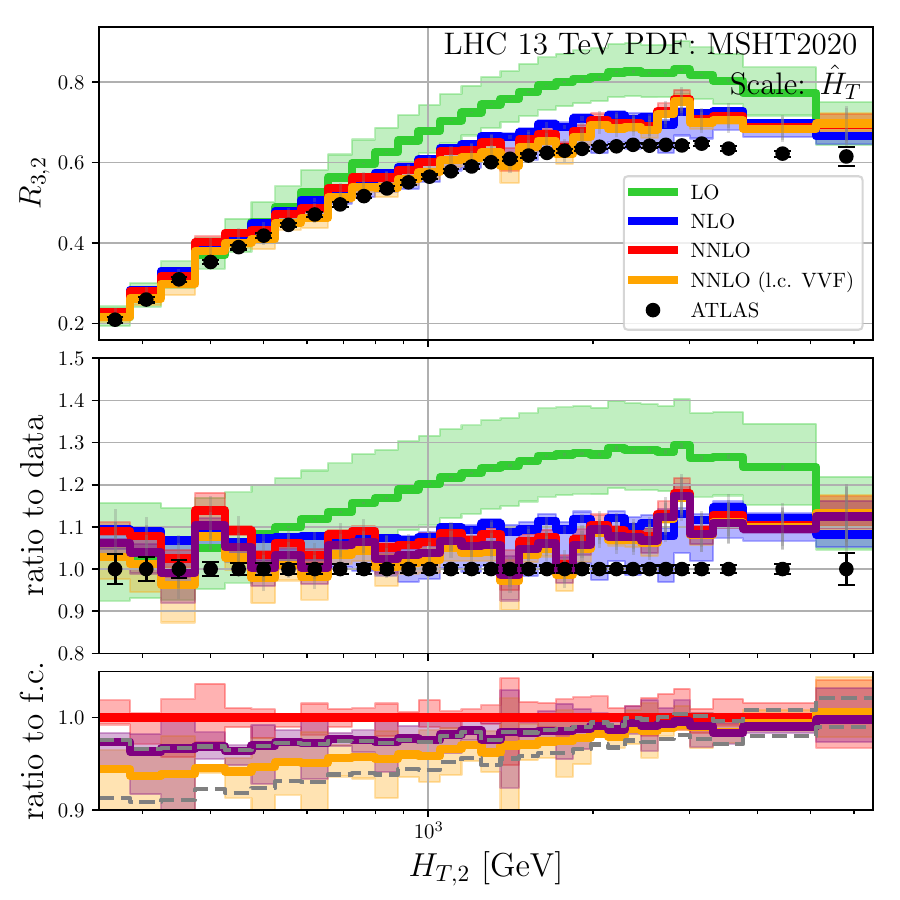}
\includegraphics[width = 0.32\textwidth,page=2]{figures/final_pp_jjj_2405.20206_MSHT2020_HThat}
\includegraphics[width = 0.32\textwidth,page=5]{figures/final_pp_jjj_2405.20206_MSHT2020_HThat}
\caption{Ratio of three- to two-jet cross sections, $R_{3/2}$. The format is the same as in Fig.~\ref{fig:aaa}. The data points are from Ref.~\cite{ATLAS:2024png}.}
\label{fig:R32}
\end{figure}

The fiducial phase space for $R_{3/2}$ at LHC @ 13TeV is defined very similarly to the one for TEEC. In particular, jets are clustered with the anti-$k_T$ algorithm with $R = 0.4$ and $p_T > 60$ GeV, but with $|y| < 4.5$ which is larger than for TEEC. The observable is binned in $H_{T,2}$, Eq.~\eqref{eq:HT2}, with $H_{T,2} > 250$ GeV, and in slices of $p_{T,3}$, the transverse momentum of the third-hardest jet, with $p_{T,3} > 60$ GeV. The theoretical predictions have been obtained with the \textsc{MSHT2020} NNLO pdf set \cite{Bailey:2020ooq} using the renormalisation and factorisation scales Eq.~\eqref{eq:ScaleHT}. The results including the size of the subleading-color effects are presented in Fig.~\ref{fig:R32}. We notice that for lower values of $H_{T,2}$ and independently of the cut on $p_{T,3}$, the difference between full-color and leading-color $\msbar$ approximate results can grow up to 5\%. Furthermore, this difference is not comprised within the scale variation. Unfortunately, inclusion of the subleading-color contributions results in a larger difference between theory and data, as the NNLO corrections become smaller compared to the NLO result. Catani's finite remainder performs noticeably better in this case and the difference to the exact result is less pronounced, at most 2\%.
\section{Prior-based estimates of subleading-color effects} \label{sec:TNP}

The functional dependence of the two-loop finite remainders on $N_c$ and $Q_f$ is known without computing the full-color amplitude. Based on this structural knowledge alone, one can attempt to estimate the size of the subleading-color contributions, and thus assign an uncertainty to a calculation based on the leading-color approximation.

Assuming that the coefficients $c_{i,j}$ in Eqs.~\eqref{eq:ppaaa_fin2}~and~\eqref{eq:ppaaj_fin2} are of similar size, we construct a prior based on the leading color result. For the pure-parton amplitudes the first subleading term is of $\order{1/N_c^{2}}$. We, therefore, parameterize the subleading-color terms in the following way.
\begin{equation}
\begin{split}
    &\frac{ 2 \, \text{Re} \, \ip*{{\cal M}^{(0)}}{{\cal F}^{(\msbar,2)}(\mu = \hat{E}_{\text{CMS}})} + \ip*{{\cal F}^{(\msbar,1)}(\mu = \hat{E}_{\text{CMS}})}{{\cal F}^{(\msbar,1)}(\mu = \hat{E}_{\text{CMS}})}  }{\ip*{{\cal M}^{(0)}}{{\cal M}^{(0)}}}\Bigg|_{\rm l.c.}^{\theta} =  \bigg(N_c^2 + \frac{\theta}{N_c^2}\bigg) \, c_{1,0}\,.
\end{split}
\end{equation}
For the amplitudes involving photons, the parametrically largest subleading-color coefficients arise from (for $n_f=5$ with two up- and three down-type quarks) $\sum_{q'} \frac{Q_{q'}^2}{Q_q^2} N_c = \frac{11}{9Q_q^2} N_c$. We, therefore, use the parameterization
\begin{equation}
\begin{split}
    &\frac{ 2 \, \text{Re} \, \ip*{{\cal M}^{(0)}}{{\cal F}^{(\msbar,2)}(\mu = \hat{E}_{\text{CMS}})} }{\ip*{{\cal M}^{(0)}}{{\cal M}^{(0)}}}\Bigg|_{\rm l.c.}^{\theta} = N_c^2 \bigg[ c_{1,0} + \frac{n_f}{N_c} c_{2,0} \bigg] + \sum_{q'} \frac{Q_{q'}^2}{Q_q^2} \, n_f \, c_{5,0} + \sum_{q'} \frac{Q_{q'}^2}{Q_q^2}N_c\theta\, c_{1,0}\,,
\end{split}
\end{equation}
for the $pp \to \gamma\gamma j$ amplitudes, and
\begin{equation}
\begin{split}
    &\frac{ 2 \, \text{Re} \, \ip*{{\cal M}^{(0)}}{{\cal F}^{(\msbar,2)}(\mu = \hat{E}_{\text{CMS}})} }{\ip*{{\cal M}^{(0)}}{{\cal M}^{(0)}}}\Bigg|_{\rm l.c.}^{\theta} = N_c^2 c_{1,0} + \sum_{q'} \frac{Q_{q'}^2}{Q_q^2}N_c\theta\, c_{1,0}\,,
\end{split}
\end{equation}
for the $pp \to \gamma\gamma\gamma$ case. The nuisance parameter $\theta$ is varied in the range $[-1,1]$ in order to obtain an uncertainty band. The estimate is shown in the lowest panel of Figs.~\ref{fig:aaa}~to~\ref{fig:R32}.

Overall, we find that this simple parameterization yields a reasonable uncertainty estimate that accounts for the differences between the $\msbar$ and Catani's leading-color finite remainder, as well as the full color-finite remainder. The only mild exception is given by the $R_{3/2}$ results, shown in Fig.~\ref{fig:R32}. At low $H_{T,2}$, we find a difference between the $\msbar$ finite remainder and the full-color results that is twice larger than our uncertainty estimate.

\section{Conclusions}

We have presented several updated results for processes with final-state photons and/or jets, in particular for $pp \to \gamma\gamma\gamma$ and $pp \to 3$ jets. The improvement consisted in including subleading-color corrections in virtual amplitudes. In most cases, the effect turned out to be small and covered by the uncertainty estimate through scale variation, even though scale variation has nothing to do with subleading color. The small size of the effect was usually due to the size of the complete double-virtual contributions rather than the reliability of the leading-color approximation.

We studied two different leading-color finite-remainder definitions, and have shown that there is in general no preference for either. Only in the case of jet cross sections, did Catani's definition perform noticeably better.

As a warning for the future, we have uncovered relatively sizeable subleading-color effects in the ratio of three- to two-jet cross sections. This case is particularly disappointing, since leading-color results showed better agreement with data.

\begin{acknowledgments}
This work was supported by the Deutsche Forschungsgemeinschaft (DFG) under grant 396021762 -
TRR 257: Particle Physics Phenomenology after the Higgs Discovery. The authors gratefully acknowledge the computing time provided to them at the NHR Center NHR4CES at RWTH Aachen University (project number p0020025). This is funded by the Federal Ministry of Education and Research, and the state governments participating on the basis of the resolutions of the GWK for national high performance computing at universities (\url{www.nhr-verein.de/unsere-partner}). This work was performed using the Cambridge Service for Data Driven Discovery (CSD3), part of which is operated by the University of Cambridge Research Computing on behalf of the STFC DiRAC HPC Facility (www.dirac.ac.uk). The DiRAC component of CSD3 was supported by STFC grants ST/P002307/1, ST/R002452/1 and ST/R00689X/1
\end{acknowledgments}

\appendix

\section{Setups for multi-photon cross sections} \label{app:setups}

\subsection*{Smooth-cone photon isolation}

The processes $pp \to \gamma\gamma\gamma$ and $pp \to \gamma\gamma+$jet require a special treatment of photons that are nearly collinear to hadrons. A consistent framework for this problem requires fragmentation functions. As an alternative, one can use a proxy proposed by Frixione \cite{Frixione:1998jh}. In this case, one considers the sum of transverse energies, $E_{Tk}$, of final-state partons $k$ separated by $\Delta R_{k\gamma} < \Delta R$ from a photon
\begin{equation}
E_T^{\rm iso}(\Delta R) \equiv \sum_{\substack{\text{final-state} \\ \text{parton } k}} E_{Tk} \, \theta (\Delta R - \Delta R_{k\gamma})\,.
\end{equation}
An event is accepted, if for any $\Delta R \le \Delta R_0$
\begin{equation}
E_T^{\rm iso}(\Delta R) < E_T^{\rm max} \frac{1-\cos(\Delta R)}{1-\cos (\Delta R_0)}\,.
\end{equation}
The parameters $\Delta R_0$ and $E_T^{\rm max}$ can be chosen freely to minimize the difference with respect to a treatment based on fragmentation functions.

\subsection*{$pp \to \gamma\gamma\gamma$}

The results of Ref.~\cite{Chawdhry:2019bji} were obtained with the {\tt NNPDF31\_nnlo\_as\_0118} pdf set \cite{NNPDF:2017mvq}, and renormalisation and factorisation scales
\begin{equation}
\mu_R = \mu_F = \frac{H_T}{4} \,, \qquad H_T \equiv \sum_{i=1}^3 p_T(\gamma_i) \,.
\end{equation}
The calculation was designed to match the specification of the experimental analysis \cite{ATLAS:2017lpx}
\begin{itemize}
\item LHC @ 8TeV
\item $p_T(\gamma_i) > p_T(\gamma_j)$ for $i < j$
\item $p_T(\gamma_1) > 27$ GeV, $p_T(\gamma_2) > 22$ GeV and $p_T(\gamma_3) > 15$ GeV,
\item $|\eta(\gamma)|< 1.37$ or $1.56 < |\eta(\gamma)| < 2.37$,
\item $\Delta R(\gamma_i,\gamma_j) > 0.45$ for $i \neq j$,
\item $m(\gamma\gamma\gamma)>50$ GeV.
\item The photons satisfy Frixione's criterion with $\Delta R_0 = 0.4$ and $E_T^{\rm max} =10~{\rm GeV}$.
\end{itemize}

\subsection*{$pp \to \gamma\gamma+$jet}

The results of Ref.~\cite{Chawdhry:2021hkp} were obtained with the {\tt NNPDF31\_nnlo\_as\_0118} pdf set \cite{NNPDF:2017mvq}, and renormalisation and factorisation scales
\begin{equation}
\mu_R=\mu_F = \frac{1}{2} \sqrt{m^2(\gamma\gamma)+ p_T^2(\gamma\gamma)} \,.
\end{equation}
At the time when Ref.~\cite{Chawdhry:2021hkp} was finalized, there were no published measurements for the considered process. On the other hand, this process is a background for Higgs-boson + jet production with Higgs-boson decay into photons. Taking this as a motivation, cuts inspired by Higgs-boson analyses \cite{ATLAS:2015zhl,CMS:2019pov} were adopted
\begin{itemize}
\item LHC @ 13TeV,
\item $p_T(\gamma_1) > p_T(\gamma_2)$,
\item $p_T(\gamma_1) > 30$ GeV and $p_T(\gamma_2) > 18$ GeV,
\item $|\eta(\gamma)| < 2.4$,
\item $\Delta R(\gamma_1,\gamma_2) > 0.4$,
\item $p_T(\gamma\gamma) > 20$ GeV,
\item $m(\gamma\gamma) > 90$ GeV,
\item The photons satisfy Frixione's criterion with $\Delta R_0 = 0.4$ and $E_T^{\rm max} =10~{\rm GeV}$.
\end{itemize}

\newpage
\bibliographystyle{JHEP}
\bibliography{main} 

\end{document}